\documentclass[aps,preprint,preprintnumbers,nofootinbib,showpacs]{revtex4}

\usepackage{graphicx,color,amsmath}
\begin{document}
\title{Hadronic production of light color-triplet Higgs bosons: an 
alternative signature for GUT}
\author{Kingman Cheung$^{1)}$ and Gi-Chol Cho$^{2)}$}
\affiliation{
$^1$
National Center for Theoretical Sciences, National Tsing Hua 
University, Hsinchu, Taiwan, R.O.C. \\
$^2$
Department of Physics, Ochanomizu University, Tokyo 112-8610, Japan}
\date{\today}

\begin{abstract}
The conventional signature for grand unified theories (GUT) is the
proton decay.  Recently, some models in extra dimensions or with
specific discrete symmetries, which aim at solving the doublet-triplet
problem, allow the color-triplet in the TeV mass region by suppressing
the Yukawa couplings of the triplets to matter fermions.  We study the
hadronic production and detection of these TeV colored Higgs bosons as
an alternative signature for GUT, which would behave like massive
stable charged particles in particle detectors producing a striking
signature of a charged track in the central tracking system and being
ionized in the outer muon chamber. We found that the LHC is sensitive
to a colored Higgs boson up to about 1.5 TeV.  If the color-triplets
are stable in cosmological time scale, they may constitute an
interesting fraction of the dark matter.
\end{abstract}
\preprint{NSC-NCTS-021205, OCHA-PP-196}
\maketitle

\section{Introduction}

The doublet-triplet splitting problem is one of the most serious 
problems in supersymmetric (SUSY) grand unified theories 
(GUT)~\cite{susy-gut}. 
In SUSY-GUT, the weak-doublet Higgs fields which are responsible 
for the electroweak symmetry breaking belong to the $\textbf{5}$ 
and $\bar{\textbf{5}}$ representations of SU(5). 
They are composed of, in addition to the weak doublet, the color-triplet 
under the gauge group of the standard model (SM). 
After the SU(5) symmetry breaking, the weak doublets would still be
massless while the color-triplets have to be decoupled 
from physics below the GUT scale, $M_{\rm GUT}\sim 10^{16}\,{\rm GeV}$, 
otherwise
they may induce the proton decay in an experimentally unacceptable 
level through the Yukawa interactions to quarks or leptons 
in the SM. 
Most attempts to the doublet-triplet problem in the literatures
have been focused on how to naturally explain the 
hierarchy between the weak doublet and the color-triplet Higgs boson 
masses after the GUT symmetry breaking~\cite{DT-splitting}. 

An alternative approach to the doublet-triplet 
splitting problem, instead of requiring the triplet to have a 
mass of GUT scale, 
is to suppress the Yukawa couplings of the color-triplet Higgs fields 
to the SM quarks or leptons in order to preserve the proton longevity.
Thus, no mass splitting between the weak-doublet and color-triplet 
Higgs fields is required.
The natural scale of the color-triplet Higgs mass in this approach 
is the electroweak scale, say $O(100\,{\rm GeV})\sim O(1\,{\rm TeV})$.
Suppression of the Yukawa couplings of the triplet is possible via
the following mechanisms.
\begin{itemize}
\item
The SM fermion masses are generated from a higher dimensional 
operators involving the GUT Higgs field in the adjoint representation,
the VEV of which can be arranged such that the Yukawa couplings of
the doublets get the usual values while the Yukawa couplings of the 
triplets are zero~\cite{Dvali:1995hp}.
\item
The overlap of the wave-functions of the color-triplet Higgs fields 
with the SM fermions in extra dimensions are sufficiently 
small~\cite{Haba:2002if}. 
\item
Another type of models is based on orbifolding in AdS space.  By assigning
different spatial parities to various components of the Higgs multiplet,
the wave-function of the color-triplet Higgs fields are zero at the
Planck brane, on which the matter fermions reside \cite{nomura}. Thus,
the excessive proton decay via the the colored Higgs boson is highly
suppressed.  In addition, the mass of the color-triplet fields is given 
by the warp factor of the AdS and naturally of a TeV scale.
\end{itemize}
We shall describe these models in some details in the next section.

The generic feature of these models to the doublet-triplet 
problem is the presence of colored Higgs bosons in TeV scale. 
In fact, this is an alternative novel signature to the conventional
signature of proton decay, and can be tested in the upcoming LHC.
It is then worthwhile to investigate the phenomenology of the TeV
colored Higgs bosons in collider experiments at high energy frontier. 
In hadronic collisions, the colored Higgs bosons are produced via 
the glue-glue fusion and the $s$-channel $q\bar{q}$ annihilation. 
On the other hand, 
it is also possible to produce pairs of colored Higgs bosons 
at TeV $e^+ e^-$ linear colliders and $\gamma \gamma$ 
colliders via the U(1)$_Y$ gauge boson exchange, but the production 
rates  are suppressed relative to hadronic production because of 
the small U(1)$_Y$ coupling. 

In this paper, we calculate the production and describe the detection of 
the TeV colored Higgs bosons in hadron colliders. 
As we already mentioned, the colored Higgs bosons do not have sizable 
Yukawa couplings to the SM fermions in order to suppress the fast 
proton decay. 
Thus, the only allowed production channels of the colored Higgs bosons
in hadronic collisions are via the SU(3)$_C$ invariant 
interactions. 
The colored Higgs bosons have a distinctive feature
that gives rise to a novel signature like a ``heavy muon''.
Since they have no Yukawa couplings and can only pair-wise couple 
to gluons,  the colored Higgs bosons cannot decay into the SM 
particles directly. 
They may be able to decay into a gluino and a colored higgsino 
through the gaugino-matter interactions only if it is kinematically 
allowed, though. 
However, in our analysis, we assume that the colored Higgs boson
has a mass less than the sum of gluino and colored higgsino masses so 
that such a decay channel is forbidden. 
This is a reasonable assumption because we expect the colored higgsino
to have a mass scale as heavy as the colored Higgs boson.
Therefore,  the colored Higgs bosons produced in hadron colliders are 
stable, and will be hadronized into color-neutral massive particles
by combining with gluons or light quarks.
Statistically, we also assume that 
half of the time the colored Higgs boson will be an
electrically-charged particle after hadronization.
\footnote{
The probability of hadronizing into a charged particle depends on the
spectrum of the bound states of the colored Higgs boson with the 
light degrees of freedom.  However, there does not exist any 
realistic calculations of the spectrum, and so we take
the probability to be $1/2$.}
This is the reason
why it will behave like a ``heavy muon'' in the detector.  
The novel signature will then be an observable charged track in the
central tracking chamber and/or the silicon vertex detector
 and a penetration to the outer muon chamber --
heavy muon. 
Taking into account of all these properties, we study the discovery 
potential of the colored Higgs bosons in the upcoming hadron colliders,
the LHC and the VLHC.

This paper is organized as follows. 
In the next section, we briefly describe the three models mentioned 
above that allow light color-triplet Higgs fields.
In Sec. III,  we review the interactions of colored Higgs 
bosons and briefly discuss the indirect constraints.
In Sec.~\ref{sec:detection}, we discuss the conditions to detect 
the colored Higgs bosons. 
The hadronic production of the colored Higgs bosons and 
their detectability in collider experiments will be discussed 
in Sec.~\ref{sec:production}. 
The last section will be devoted to a summary and discussion. 
\section{Models of TeV color-triplet}

In this section, we highlight on three models that allow light (TeV) 
color-triplet Higgs bosons by Dvali \cite{Dvali:1995hp}, by 
Haba and Maru \cite{Haba:2002if}, and by
Goldberger, Nomura and Smith \cite{nomura}.

In supersymmetric GUT models, the Higgs doublets that give Yukawa 
couplings to fermions are accompanied by the color-triplet in the same 
multiplet. 
The color-triplet couples to fermions with the same Yukawa couplings 
as the doublet before the GUT symmetry is broken.  That is why the 
color-triplet has to be extremely heavy in order to avoid the proton decay.  
However, the proton decay problem can also be solved if the Yukawa 
couplings of the color-triplet are suppressed relative to the doublet 
by a ratio $M_W/M_{\rm GUT}$, and thus the color-triplet needs not to 
be heavy and can be as light as the doublet. 
Imposing a specific discrete symmetry to forbid the lowest order 
Yukawa term
${\textbf{16}}^\alpha {\textbf{16}}^\beta {\textbf{10}}$, the Yukawa 
couplings of the doublet have to be generated via higher dimensional 
terms. Dvali \cite{Dvali:1995hp} constructed a higher dimensional 
SO(10) invariant operator
\[
\frac{Y_{\alpha \beta}}{M} \, {\textbf{10}}_i \, {\textbf{45}}_{ik} \, 
{\textbf{16}}^\alpha\,  \gamma_{k} \, {\textbf{16}}^\beta
\]
where $M \sim M_{\rm GUT}$, 
the matter fermions reside in the $\textbf{16}$, $\alpha,\beta$ are 
family indices, $\textbf{10}_i$ consists of color-triplets ($i=1,...,6$) 
and doublets ($i=7,...,10$), and $\textbf{45}$ is the GUT Higgs in the
adjoint representation of SO(10).  This term can arise from tree-level 
exchanges of heavier states at $M_{\rm GUT}$.  Taking a VEV for the 
$\textbf{45}$ as 
\[
\langle {\textbf{45}}_{ik} \rangle = diag(0,0,0,A,A) \otimes \epsilon
\]
where $A\sim M_{\rm GUT}$ and $\epsilon$ is the $2\times 2$ 
antisymmetric matrix, and  substituting into the above operator, 
it is easy to see that the
doublet gets a Yukawa coupling $Y_{\alpha\beta} A/M$ while the 
color-triplet has zero Yukawa couplings.  In this case, the 
color-triplet would not cause any proton decay and thus can be as light 
as the doublet. 
Of course, the usual ${\textbf{16}}^\alpha {\textbf{16}}^\beta 
{\textbf{10}}$ Yukawa term is absent because of a specific discrete 
symmetry. Dvali was using a $Z_2 \times Z_3$ \cite{Dvali:1995hp}.

Another interesting solution to the doublet-triplet problem was
recently proposed by Haba and Maru \cite{Haba:2002if}.  Although it is 
of similar spirit, the setup is however entirely different.
Basically, in the setup of extra dimensions the proton stability is
maintained by suppressing the Yukawa couplings of the color-triplet to 
matter fermions through the small overlap of wave-functions in the 
extra dimensions.  They started with a SU(5) model in 5 dimensions 
with the Higgs doublets and triplets in ${\textbf 5}$ and $\bar {\textbf
5}$, the usual SU(5) GUT adjoint Higgs field being assumed to 
localize on the brane at $y=0$ ($y$ is the coordinate in the extra
dimension), and a pair of bulk fields in  
${\textbf{24}}$.  
Through interactions with the bulk fields (very similar to the 
idea of domain-wall fermions), the zeroth mode of the Higgs doublets
are localized in a position close to the brane at $y=0$, where the
matter fermions are confined, while the color-triplet Higgs fields are
localized at a further distance from the brane at $y=0$.  By varying
the distance between the doublets and triplets, the Yukawa couplings
of the triplets can be sufficiently suppressed to avoid the proton
decay while the doublets can generate the usual Yukawa couplings to
the matter fermions.  Therefore, the color-triplets need 
not be heavy and can be as light as the weak doublets.

The third type of models is based on orbifolding. By assignment of
different spatial parities (or boundary conditions) to various
components of a multiplet, the component fields can have very
different properties at the fixed points.  Thus, it is possible to
break a symmetry or to achieve the doublet-triplet splitting by the boundary
conditions.
In the model by Goldberger et al. \cite{nomura}, they started from
the Randall-Sundrum scenario \cite{randall}: a slice of AdS space
with two branes (the Planck brane and the TeV brane) at both ends.
The extra dimension is compactified on an $S^1/Z_2$ orbifold.  The
hierarchy of scales is generated by the AdS warp factor $k$, which is
of order of the five-dimensional Planck scale $M_5$, such that the 4D
Planck scale is given by $M^2_{\rm Pl} \sim M_5^3/k$.  The fundamental
scale on the Planck brane is $M_{\rm Pl}$ while the fundamental scale
on the TeV brane is rescaled to TeV by the warp factor: $T\equiv k
e^{-\pi k R}$, where $R$ is the size of the extra dimension.
The model is a 5D supersymmetric
SU(5) gauge theory compactified on the orbifold $S^1/Z_2$ in the AdS
space.  The boundary conditions break the SU(5) symmetry and provide
a natural mechanism for the Higgs doublet-triplet splitting and
suppress the proton decay.  The Planck brane respects the SM gauge
symmetry while the TeV brane respects the SU(5) symmetry.  The matter
fermions reside on the Planck brane.  
By the boundary conditions the
wave-function of the color-triplet Higgs fields are automatically zero
at the Planck brane, on which the matter fermions reside, while the
doublet Higgs fields are nonzero at the Planck brane and give Yukawa
couplings to the matter fermions.  Thus, the excessive proton decay
via the color-triplet Higgs fields is highly suppressed, and the
doublet-triplet splitting is therefore natural by the boundary conditions.
The mass of the color-triplet fields (and the XY guage
bosons) is given by the warp factor and is of a TeV scale, the same as
the KK states of other fields in the setup.
This model is very similar, in spirit, to the second model in
suppressing the proton decay by a very small or zero overlap of the
wave-functions, but the mechanism is more natural in this model.  In
addition, the mass of the color-triplet is also naturally given by the
warp factor in the TeV scale.

\section{Review of light colored Higgs boson}

\begin{figure}[ht!]
\begin{center} 
\includegraphics[width=4.5cm,clip]{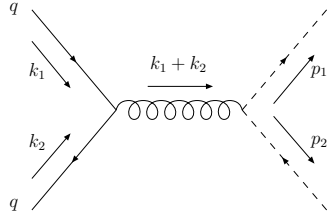}
\caption{Feynman diagram for $q\bar{q} \to H_C H_C^*$. 
}
\label{fig:qq}
\end{center}
\end{figure}

\begin{figure}[ht!]
\begin{center} 
\includegraphics[width=12cm,clip]{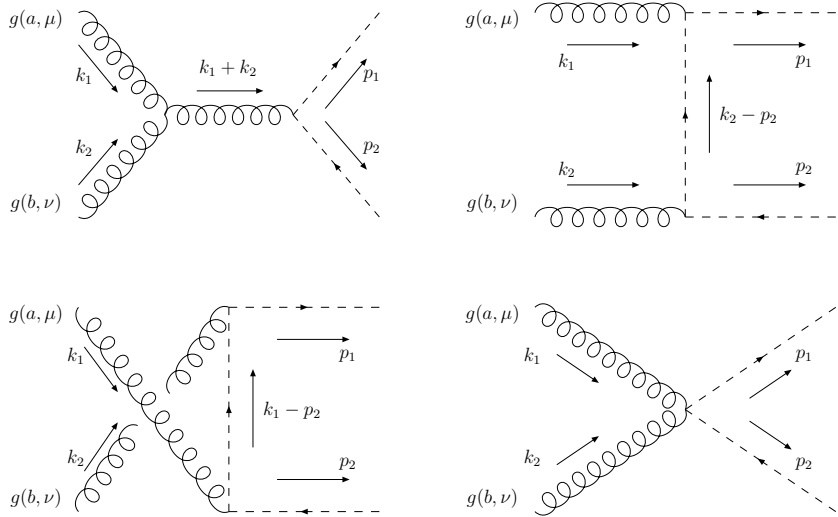}
\caption{Feynman diagrams for $gg \to H_C H_C^*$. 
}
\label{fig:gg}
\end{center}
\end{figure}

The two Higgs fields in the SUSY-GUT, $H(\textbf{5})$ and 
$H(\bar{\textbf{5}})$, are composed of the weak doublet and 
the color triplet as follows: 
\begin{eqnarray}
H(\textbf{5}) &=& (H_C, H_u), 
\\
H(\bar{\textbf{5}}) &=& (\bar{H}_C, H_d), 
\end{eqnarray}
where the weak doublets $H_u$ and $H_d$ are responsible for 
the up- and the down-type quark (lepton) masses, respectively. 
The color-triplets $H_C$ and $\bar{H}_C$ could be as light as $O(\,{\rm TeV})$ 
via the mechanisms that we have described above.
The price for the TeV colored Higgs bosons is the absence of 
the Yukawa interactions of $H_C$ and $\bar{H}_C$ to the SM fermions. 
The colored Higgs boson couples to the gluon $A_\mu^a$ 
through the following interactions: 
\begin{eqnarray}
{\cal L} &=& -i g_s H_C^* {\stackrel{\leftrightarrow}{\partial}}_\mu 
H_C T^a A^{a\mu} 
     + g_s^2 T^a T^b H_C^* H_C A^a_\mu A^{b\mu}, 
\label{int}
\end{eqnarray} 
where $T^a$ is the generator of SU(3), and 
$A {\stackrel{\leftrightarrow}{\partial}}_\mu B 
\equiv A(\partial_\mu B) - (\partial_\mu A)B$. 
The interactions for $\bar{H}_C$ is the same as $H_C$ in Eq.~(\ref{int}). 
The production of the colored Higgs bosons in the lowest order 
is via the $s$-channel $q\bar{q}$ annihilation 
and the glue-glue fusion, shown by the Feynman diagrams
in Figs. \ref{fig:qq} and \ref{fig:gg}. 

In addition to Eq.~(\ref{int}) the colored Higgs boson also 
interacts with the colored higgsino $\widetilde{h}_C$ and 
the gluino $\widetilde{g}$:  
\begin{eqnarray}
{\cal L} &=& -\sqrt{2} g_s \left(
H_C^* \, \overline{\widetilde{g}^a} \,T^a \,
\widetilde{h}_C + 
\overline{\widetilde{h}_C} \,
\widetilde{g}^a \, T^a \, H_C
\right).
\label{hhg}
\end{eqnarray} 
However, as we already mentioned, we have assumed that the colored Higgs
bosons do not decay into colored higgsinos and gluinos because it
is not allowed kinematically.

Is there any indirect constraints on the colored Higgs bosons from 
the high-energy experiments? 
Let us consider the $Z$-pole experiments at LEP1 and SLC 
where the radiative corrections to the gauge boson propagators 
and the $Z \to f\bar{f}$ ($f$ denotes quarks or leptons) 
vertices are severely constrained. 
The contributions to the gauge boson propagators are summarized 
by the $S,T,U$ parameters~\cite{Peskin}. 
It is well known that the SU(2)$_L$ singlet scalars do not contribute 
to the $S,T,U$ parameters~\cite{Cho:1999km} such that the colored Higgs 
boson is free from the constraints.
Since the Yukawa interactions of $H_C$ and $\bar{H}_C$ to the quarks and 
leptons are highly suppressed, they do not contribute to the 
$Z\to f\bar{f}$ processes.
By the same reason, there is no constraint on the colored Higgs 
bosons from flavor physics experiments. 
We, therefore, conclude that no indirect constraints are implied
for the colored Higgs mass from current experiments. 

We comment on the gauge coupling unification before closing 
this section. 
It is well known that the success of gauge coupling unification 
in the minimal supersymmetric standard model (MSSM) could be preserved 
if complete multiplets of SU(5) (e.g.,  $\textbf{5}$ or $\textbf{10} 
\cdots$) are added to the spectrum of MSSM. 
When $H_C$ and $\bar{H}_C$ are in TeV scale, there must be a vector pair 
of weak doublets in the same scale to form
the $\textbf{5}$ and $\bar{\textbf{5}}$ multiplets so that the gauge 
coupling unification is unaltered. 
The origin of the weak vector-like pair is discussed 
in both Refs.~\cite{Dvali:1995hp} and \cite{Haba:2002if}.

\section{Detection of Massive Charged Particles}
\label{sec:detection}

The colored Higgs boson will hadronize into a massive stable particle,
which is electrically either neutral or charged.  
Both the neutral and charged
particles will undergo hadronic interactions with detector
materials while the charged particle will also undergo ionization, through
which the particle loses its kinetic energy (K.E.).  Such a massive stable
particle will have a high transverse momentum and a small velocity 
(or $\beta=v/c$).  Furthermore, if it is 
charged it will penetrate detector materials like a muon.

First, we discuss the hadronic energy loss of a massive stable particle
in the detector (both neutral and charged should have similar behavior.)
Although it is strongly interacting, it penetrates the material with a very 
small loss of energy.  This is because the energy 
loss in hadronic elastic scattering is 
negligible because of the small momentum transfer.  In addition, the 
energy loss in elastic or inelastic scattering with nucleons is also small
because of the huge mass difference between the massive particle (TeV) and the 
nucleon.  
%
%
Thus, the energy loss via hadronic collisions does not lead to detection
of the massive particle.  If the colored Higgs hadronizes into a
neutral particle, it will escape the detector and undetected.  The detection
of colored Higgs bosons has to rely upon the charged state, which
will be described next.

\begin{figure}[th!]
\includegraphics[height=4in,angle=270]{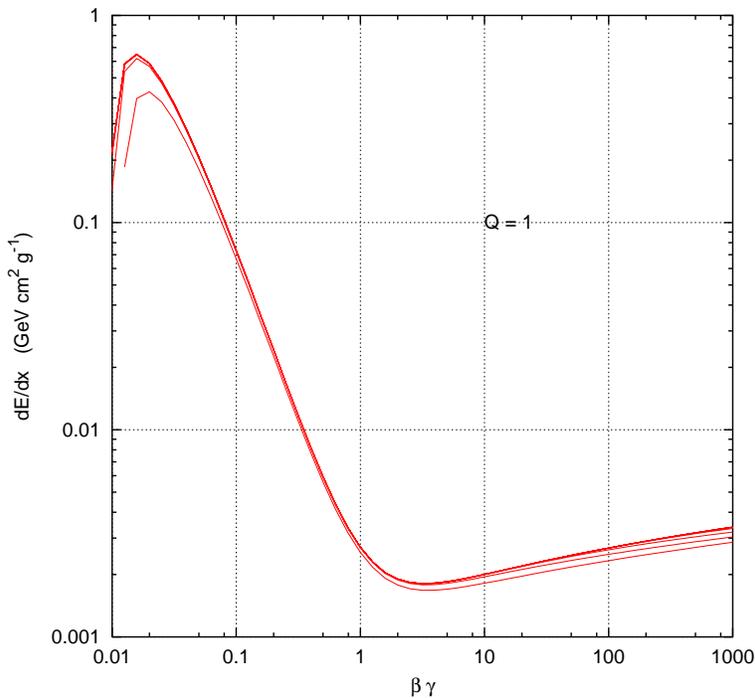}
\caption{\small \label{dedx}
$dE/dx$ calculated using the formula in PDB for a range of $M=10^{-3} - 10^4$
GeV. $M=10^{-3}$ GeV is for the bottom of the lines and goes up to 
$M=10^4$ GeV for the top of the lines.  
It is for Si (we used a value of $I=2$ eV for illustration.)
}
\end{figure}

The energy loss $dE/dx$ 
due to ionization with the detector material is very standard
\cite{pdg}. Essentially, the penetrating particle loses energy by exciting
the electrons of the material.
The basic formula can be found in Eq. (26.1) in Particle Data Book (PDB)
 \cite{pdg}.
We shall ignore the small correction from the density effect.
Ionization energy loss $dE/dx$ 
is a function of $\beta \gamma$ and the charge $Q$ of the
penetrating particle.  The dependence on the
mass $M$ of the penetrating particle comes in through $\beta\gamma$ for 
a large mass $M$ and small $\gamma$ \cite{pdg}.
In other words, 
 $dE/dx$ is the same for different masses if the $\beta\gamma$ and $\beta$ 
values of these particles are the same.  
In Fig. \ref{dedx}, we show $dE/dx$ for a wide mass range 
$10^{-3} - 10^4$ GeV as a function of $\beta\gamma$.
  For the range of $\beta\gamma$ between 0.1 and 1 
that we are interested in, $dE/dx$ has almost 
no explicit dependence on the mass $M$
of the penetrating particle. 
Therefore, when $dE/dx$ is measured in an experiment, the $\beta\gamma$ can
be deduced, which then gives the mass of the particle if the momentum $p$ is
also measured.  Hence, $dE/dx$ is a good tool for particle identification
for massive stable charged particles. 

In fact, the CDF Collaboration did a few searches for massive stable 
charged particles \cite{cdf1,cdf2,cdf3}. 
The CDF analyses require 
that the particle produces a track in the central tracking chamber and/or
the silicon vertex detector,
and at the same time penetrates to the outer muon chamber.
The CDF detector has a central
tracking chamber and a silicon vertex detector (the central tracking
chamber has a slightly better resolution in this regard), 
which can measure the energy loss ($dE/dx$) 
of a particle via ionization, especially at low $\beta\gamma < 0.85$ 
($\beta< 0.65$) where
$dE/dx \sim 1/\beta^2$. Once the $dE/dx$ is measured, 
the mass $M$ of the particle can be determined if the momentum $p$ is
measured simultaneously.
Furthermore,  the particle is required to  
penetrate through the detector material and make
it to the outer muon chamber,
provided that it has an initial $\beta >0.25-0.45$ depending on the mass of 
the particle \cite{cdf1}.  Therefore, the CDF requirement on 
$\beta$ or $\beta\gamma$ is
\begin{equation}
\label{cut}
0.25-0.45 \;\; \alt \;\;\; \beta  \;\;\; < \;\; 0.65  \;\;\;
\Leftrightarrow \;\;\;
0.26-0.50 \;\; \alt \;\;\; \beta \gamma \;\;\; < \;\; 0.86  \;.
\end{equation}
The lower limit is to make sure that the penetrating particle can make it to 
the outer muon chamber while the upper limit makes sure that the ionization
loss in the tracking chamber is sufficient for a detection.  
We shall employ a similar requirement in our analysis that we shall illustrate
next.

\begin{figure}[th!]
\includegraphics[height=4in,angle=270]{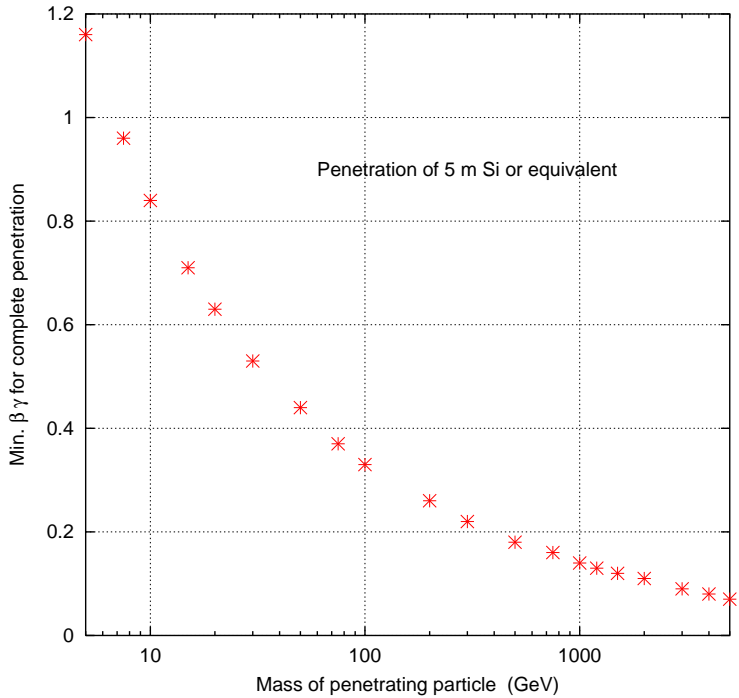}
\caption{\small \label{min-betagamma}
Minimum $\beta\gamma$ values for a complete penetration of 
a 5 m Si or equivalent vs the mass $M$ of the penetrating particle.
}
\end{figure}

Here we verify that the lower limit on $\beta\gamma$ is valid
for a 1 TeV charged particle with $Q=1$.  
We use the $dE/dx$ formula in PDB for 
Si (we choose the value $I=2$ eV for the illustration purpose.)
We calculate the minimum value of $\beta \gamma$ for a singly-charged particle
of mass $M$ to penetrate a silicon of a 5 m thickness (it is roughly
equivalent to a 1.5 m of Fe because the density of Fe is 3.4 times of that of
Si.)  The calculation procedures are rather straightforward and described as
follows.
(i) The particle of mass $M$ starts with an initial $\beta\gamma$. 
(ii) Calculate the value of $dE/dx$ and let the particle penetrate for a 
step of 1 cm. 
(iii) Evaluate the remaining energy and calculate the corresponding 
$\beta\gamma$, which is then input to calculate the next $dE/dx$ in the next 
step (1 cm).  
(iv) Repeat until the $\beta\gamma$ goes to zero or the particle
penetrates a distance of 5 m.  If $\beta\gamma$ goes to zero before 
reaching 5 m, the particle does not have a complete penetration, else 
the particle completes the penetration.  
We show the minimum $\beta\gamma$ values for a complete penetration of 
a 5 m Si or equivalent vs the mass $M$ of the penetrating particle in
Fig. \ref{min-betagamma}.  From the figure, the minimun $\beta\gamma$ for
a particle of mass between $50-500$ GeV is about $0.18-0.44$, which
is consistent with the CDF requirement in Eq. (\ref{cut}).
For a 1 TeV particle, the $(\beta\gamma)_{\rm min}$ is about $0.14$.
Therefore, in our analysis we shall choose a conservative value of
$(\beta\gamma)_{\rm min}=0.25$, and our final cut on $\beta\gamma$ is
\begin{equation}
\label{ourcut}
0.25 \;\; \le \;\;\; \beta \gamma \;\;\; \le \;\; 0.85  \;.
\end{equation}
Here we have chosen the upper limit very close to what CDF used 
\cite{cdf1,cdf2,cdf3}, because $dE/dx$ is almost independent of the mass
of the penetrating particle in the range $0.1<\beta\gamma<1$ (see 
Fig. \ref{dedx}).  
Essentially, the upper limit ensures that the particle deposits
a sufficient amount of energy in the central tracking system and the muon
system for detection, beyond which the $dE/dx$ may be too small for 
detection.

Since the lower cut on $\beta\gamma=p/M$ is $0.25$, the momentum cut is
$250$ GeV for a 1 TeV particle.  Such a cut on momentum already
makes it background free from $\mu^\pm, \pi^\pm, K^\pm$ 
together with the cut on $\beta\gamma$.  
Another configuration cut due to the detector (both CMS and Atlas) is 
\begin{equation}
\label{eta}
|\eta| < 2.5 \;.
\end{equation}
We also assume an efficiency of 80\% for each massive stable charged 
particle to be detected by the central
tracking system and the outer muon system.
This efficiency is in addition to the cuts on $\eta$ and $\beta\gamma$.

There were also some theoretical studies on search for massive stable 
charged particles, such as a light gluino as the LSP \cite{Baer:1998pg} and
a scalar tau in the context of gauge mediated supersymmetry breaking models
\cite{Feng:1997zr}.

\section{Production at hadronic supercolliders}
\label{sec:production}

Pairs of colored Higgs bosons can be produced via $q\bar q$ and
$gg$ fusion.  The color- and spin-averaged amplitude squared are given by
\begin{eqnarray}
\overline{\sum} \left | M(gg \to H_C H_C^*) \right|^2 &=& 
 \frac{g_s^4}{128} \,
\left [ 24 \left( 1 - \frac{2 \hat t_m \hat u_m}{\hat s^2} \right ) - 
\frac{8}{3} \right ]\;
\left [  1 - \frac{2 \hat s m_{H_C}^2 }{\hat u_m \hat t_m} \, \left( 1 - 
          \frac{ \hat s m_{H_C}^2 }{\hat u_m \hat t_m} \right ) \right ] \;,
 \\
\overline{\sum} \left | M(q \bar{q} \to H_C H_C^*) \right|^2 &=& 
\frac{4 g_s^4}{9} \frac{\hat t \hat u - m_{H_C}^4}{\hat s^2} \;.
\end{eqnarray}
In the above equations,  we have defined
\begin{eqnarray}
\hat t_m &\equiv& \hat t - m_{H_C}^2 = - \frac{\hat s}{2}\,(1-\beta \cos\theta)
\\
\hat u_m &\equiv& \hat u - m_{H_C}^2 = -\frac{\hat s}{2}\,(1+\beta \cos\theta),
\end{eqnarray}
where $\theta$ is the scattering angle and 
$\beta = \sqrt{1- 4m_{H_C}^2/\hat s}$ is the velocity of the outgoing
Higgs bosons in the center-of-mass frame of the incoming partons.

The parton-level differential cross sections are then given by: 
\begin{eqnarray}
\frac{d\hat \sigma}{d\cos\theta} (gg \to H_C H_C^*)  &=& 
\frac{\pi \alpha_s^2 \beta }{256 {\hat s}}  \;
\left [ 24 \left( 1 - \frac{2 \hat t_m \hat u_m}{\hat s^2} \right ) - 
\frac{8}{3} \right ]\;
\left [  1 - \frac{2 \hat s m_{H_C}^2 }{\hat u_m \hat t_m} \, \left( 1 - 
          \frac{ \hat s m_{H_C}^2 }{\hat u_m \hat t_m} \right ) \right ] \\
\frac{d\hat \sigma}{d\cos\theta} (q \bar q \to H_C H_C^*)  &=& 
\frac{2 \pi \alpha_s^2 \beta }{9 {\hat s}}  \;
  \frac{\hat t \hat u - m_{H_C}^4}{\hat s^2} 
\end{eqnarray}
After integrating over the angle $\theta$, the parton-level cross sections 
are given by
\begin{eqnarray}
\hat{\sigma}(g g  \to H_C H_C^*) &=& 
 \frac{\pi \alpha_s^2}{\hat s} \left(
\beta \frac{5 \hat s + 62m_{H_C}^2}{48\hat s} + \frac{m_{H_C}^2}{6\hat s}
\frac{4\hat s + m_{H_C}^2}{\hat s}\log \frac{1-\beta}{1+\beta}
\right),  \label{gg}  \\
\hat{\sigma}(q \bar{q} \to H_C H_C^*) &=& 
  \frac{\pi \alpha_s^2 \beta }{\hat s} \,
  \left(\frac{2}{27}-\frac{8}{27}\frac{m_{H_C}^2}{\hat s}
     \right).  \label{qq}
\end{eqnarray}
The above results agree with the cross sections
of squark-pair production in Ref.~\cite{Beenakker:1996ch} if
only the $s$-channel process is taken into account in the $q\bar q$ 
annihilation.
Note that the expressions for the production cross sections of
the $\bar H_C \bar H_C^*$ pair are the same as the $H_C H_C^*$ pair.
If the mass of $\bar H_C$ is the same as $H_C$, the sum of the cross 
sections would be doubled.  
In the minimal SUSY SU(5), they have exactly the same mass. 
Even beyond the minimal model, since there is no particular reason why
their masses should be very different, we simply take them to be equal 
and the results present in the following take into account both $H_C$ 
and $\bar H_C$.

In the calculation, we employ the parton distribution function of CTEQ v.5
(set L) \cite{Lai:1999wy} and the 
one-loop renormalized running strong coupling
constant with $\alpha_s (M_Z) =0.119$.  
The total cross section at a center-of-mass energy $\sqrt{s}$ 
is obtained by convoluting the partonic cross 
sections in Eqs. (\ref{gg}) and (\ref{qq}) with the parton 
distribution functions:
\begin{equation}
\sigma(s) = \int_{4m_{H_C}^2/s}^{1}\; d \tau \;\int_{\tau}^{1} \;\frac{dx}{x}\;
f(\tau/x) \,f(x) \; \hat \sigma( \hat s) \;,
\end{equation}
where $\hat s = \tau s$ is square of the 
center-of-mass energy of the parton-parton scattering.

\begin{figure}[th!]
\includegraphics[width=5in]{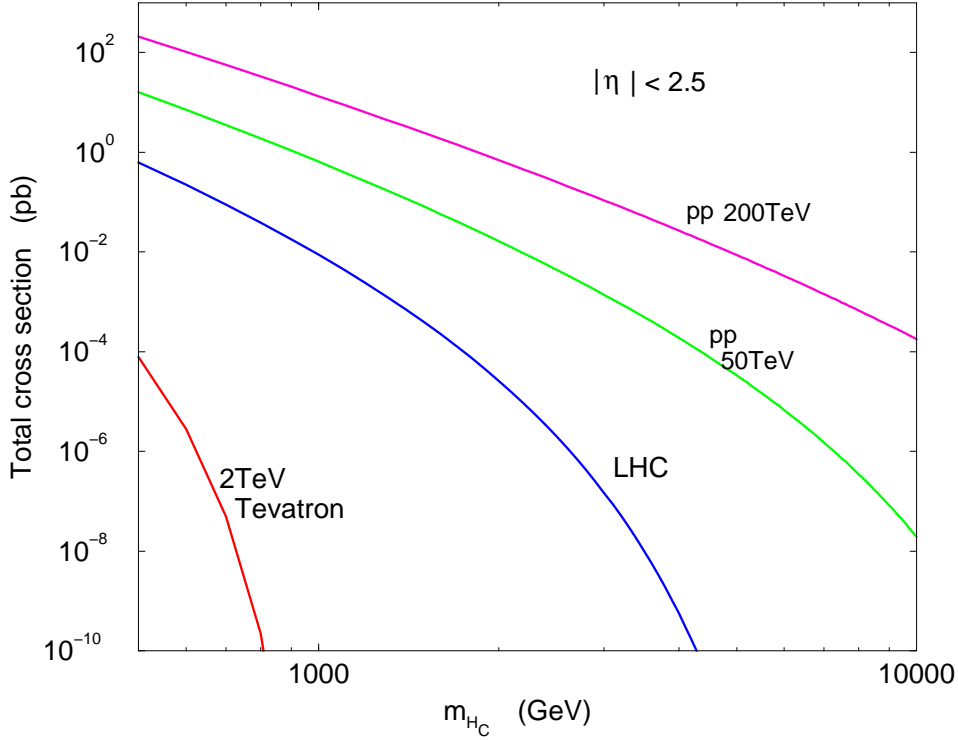}
\caption{\small\label{total}
Total cross sections for the production of the colored Higgs boson pair at
the Tevatron, LHC, and $pp$ collisions at 50 and 200 TeV. A pseudorapidity
cut $|\eta|<2.5$ is applied.
}
\end{figure}

We show in Fig. \ref{total} the production cross sections vs 
the mass $m_{H_C}$ of the 
colored Higgs boson at the 2 TeV Tevatron ($p\bar p$ collisions), at 
the LHC ($pp$ collisions), and in $pp$ collisions at $\sqrt{s}=50,200$ TeV.
The acceptance cut on the pseudo-rapidity in 
Eq. (\ref{eta}) has been applied.  As explained above this acceptance is due 
to the coverage of the central tracking chamber.  
We have shown the cross sections at 50 and 200 TeV center-of-mass energies,
which is, respectively, the lower and upper energy range of the VLHC. 
The VLHC (very large hadron collider) is another $pp$ accelerator
under discussions \cite{vlhc} in the Snowmass 2001 \cite{snowmass}.
The preliminary plan is to have an initial stage of about 40--60
TeV center-of-mass energy, and later an increase up to 200 TeV.  
The targeted luminosity is $(1-2)\times 10^{34}\,{\rm cm}^{-2} {\rm s}^{-1}$.

\begin{figure}[th!]
\includegraphics[width=5in]{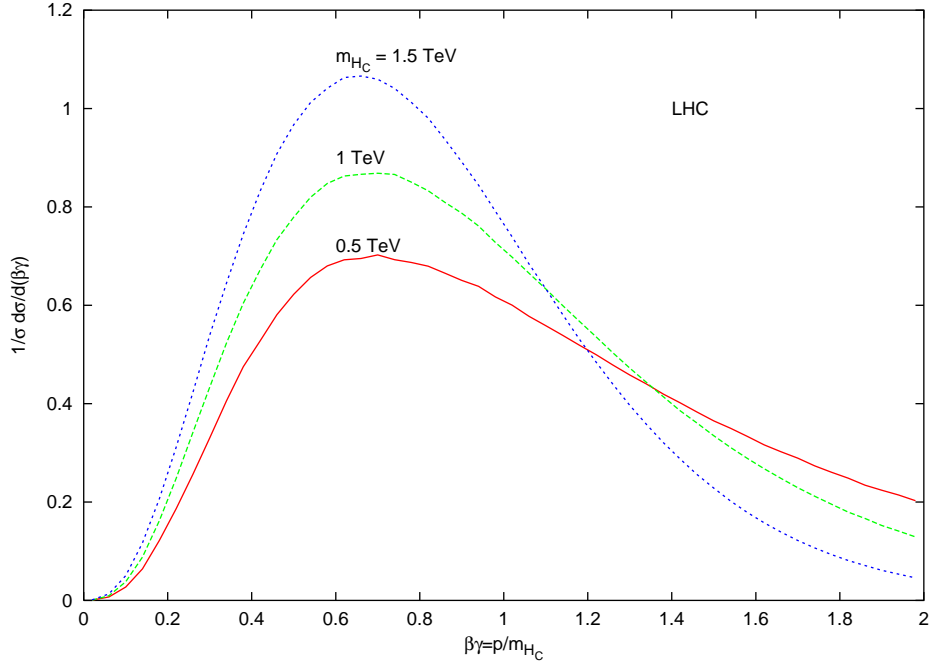}
\caption{\small\label{dist}
Normalized differential cross section $1/\sigma d\sigma/d(\beta\gamma)$ 
vs $\beta\gamma=p/m_{H_C}$ for $m_{H_C}=0.5,\, 1,\, 1.5$ TeV at the LHC.
}
\end{figure}

The next important distribution in our analysis is the $\beta\gamma =p/M$
distribution. 
We show the normalized differential cross section 
$1/\sigma d\sigma/d(\beta\gamma)$
vs $\beta\gamma$ for $m_{H_C}=0.5,\,1,\,1.5$ TeV at the LHC 
in Fig. \ref{dist}.  The majority of the cross section is below $\beta\gamma
 \simeq 1.2$ as the heavy colored Higgs bosons are produced not too
far away from the threshold.  Thus, the velocity of the Higgs bosons
is not too large.  It is obvious that the heavier the boson, the smaller is
the average velocity $\beta$.  We expect that 
when we apply the selection cut of Eq. (\ref{ourcut}) on the
colored Higgs bosons, a majority of the cross section remains.
We have verified that if we require at least one of the colored
Higgs bosons for $m_{H_C}=1$ 
TeV satisfying the $\beta\gamma$ cut of Eq. (\ref{ourcut}), 
about 60\% of the cross section remains.

Since we have assumed the detection efficiency of a track is 80\% in
addition to the acceptance cuts on $\eta$ and $\beta\gamma$,
the combined efficiency to see two tracks would be $(0.8)^2=0.64$.
In order to increase the efficiency we require to see only one of them, and
the efficiency to detect at least one of them is then $0.96$.
Therefore, if there is only one Higgs boson satisfying the $\beta\gamma$ 
cut in the final state, the detection efficiency would be 80\%.
While there are two Higgs bosons satisfying the $\beta\gamma$ 
cut in the final state, the detection efficiency for at least one of them 
would be 96\%.
Thus, the overall efficiency is more than 50\%, which is sufficient for
a sizable cross section.
We show the final cross sections for various energies and mass $m_{H_C}$ 
with the selection cuts applied and efficiencies in Table \ref{table1}.
The number of observed events shown in Table \ref{table1} includes the
following factors:
\begin{itemize}
\item[(i)]
 a probability of $1/2$ that a colored Higgs boson will hadronize into 
an electrically charged particle;
\item[(ii)]
requiring at least one of the colored Higgs bosons satisfying the selection 
cuts on $\eta$ of Eq. (\ref{eta}) and on $\beta\gamma$ of Eq. (\ref{ourcut});
\item[(iii)]
an efficiency factor of $0.8$ for each detected track (requiring to detect
at least one track), and
\item[(iv)]
both channels of $H_C H_C^*$ and $\bar H_C \bar H_C^*$ production.
\end{itemize}

Since the search is background free, the discovery or evidence of existence
for the colored Higgs bosons depends crucially on the number of observed
events, which we choose to be 10 events.  The run II with an integrated
luminosity of 20 fb$^{-1}$ at the Tevatron is sensitive up to about 400
GeV while the LHC with an integrated luminosity of 100 fb$^{-1}$ can probe
up to about 1.5 TeV.  The VLHC running at 50 TeV and 200 TeV is sensitive
to a colored Higgs boson of mass up to 3.5 and 9 TeV, respectively.

\section{Summary}

The presence of light color-triplet Higgs fields in TeV mass scale is
an alternative signature for GUT, instead of proton decay.  This is 
made possible through some mechanisms to suppress the Yukawa couplings
of the triplets to the matter fermions.  We have reviewed three possible
models by Dvali \cite{Dvali:1995hp}, by Haba and Maru \cite{Haba:2002if},
and by Goldberger, Nomura and Smith \cite{nomura}.
The former used a discrete symmetry while the latter two used a setup
in extra dimensions to achieve the suppression.  

The novel signature of these TeV colored Higgs bosons would be like
massive stable charged particles, ``heavy muons'', producing a 
track in the central tracking chamber and penetrating to the outer muon 
system.  Such a signature is background free and the momentum can
be measured, which enables determination of the mass
of the particle via the ionization-energy loss spectrum.
We have demonstrated in details that we employed a reasonably 
conservative requirement on $\beta\gamma=p/M$ cut on the massive charged
particle such that it can penetrate to the outer muon system and 
produce a charged track in the central tracking chamber.

We have calculated the production cross sections and the number of
signal events of the colored Higgs bosons.  We have taken into account
the fact that only half of the colored Higgs bosons would hadronize into
charged particles and an efficiency of 80\% to detect a track.  The number
of observable events is increased by relaxing the requirement to seeing
both colored Higgs bosons to seeing at least one.  The Tevatron Run IIb
may be able to discover a colored Higgs boson up to about 400 GeV if
an order of 10 events is required for discovery.  The LHC with an
accumulated luminosity of 100 fb$^{-1}$ is sensitive to about 1.5 TeV.
The VLHC running at 50 and 200 TeV is sensitive up to 3.5 and 9 TeV,
respectively.

A few more comments are in order.
\begin{itemize}
\item[(i)]
In this work, we have assumed that the colored Higgs bosons are stable.
An alternative is that the colored Higgs boson decays into a gluino and 
a colored higgsino if kinematically allowed.  The colored higgsino is the
supersymmetric partner of the colored Higgs boson.  This decay would also
be phenomenologically interesting because the colored higgsino would be
likely to hadronize
into a massive stable charged particle because there are no other lighter
particles that it can decay to, and the gluino would decay into quarks
and squarks producing jets and missing energies. 

\item[(ii)]
Since the colored Higgs bosons are stable over cosmological time scale,
they have a relic density since the early universe.  Our preliminary 
estimate of the relic density of $H_C$ 
is of order of $\Omega_{H_C} h^2  \sim 0.01
-0.05$ for $m_{H_C}=1-2$ TeV \cite{future}, 
which is an interesting fraction of the
observed cold dark matter density $\Omega_{\rm CDM} h^2 = 0.12 \pm 0.04$
at 95 \% C.L. \cite{Melchiorri:2002sw}.

\end{itemize}

\section*{Acknowledgments} 
G.C.C. thanks NCTS for warm hospitality during his visit. 
He is also grateful to N. Haba for discussion. 
The work of G.C.C. is supported in part by the Grant-in-Aid for 
Science Research, Ministry of Education, Science and Culture, 
Japan (No.13740149). K.C. is supported the National Center for Theoretical
Sciences under a grant from the National Science Council of Taiwan R.O.C.

\begin{table}[ht!]
\caption{\label{table1} \small
The number of signal events of massive stable charged particles due to
the pair production of colored Higgs bosons $H_C H_C^*$ and
$\bar H_C \bar H_C^*$ at the Tevatron, the LHC, and the VLHC of 50 and 
200 TeV.
We have already taken into account (i)
a probability of $1/2$ that a colored Higgs boson will hadronize into 
an electrically charged particle,
(ii) at least one of the colored Higgs bosons satisfying the selection 
cuts on $\eta$ and $\beta\gamma$, and
(iii)  an efficiency factor of $0.8$ for each detected track.
}
\medskip
\begin{ruledtabular}
\begin{tabular}{ccccc}
$m_{H_C}$ (TeV) & Tevatron  & LHC   & VLHC 50 TeV & VLHC 200 TeV\\
   & (${\cal L}=20$ fb$^{-1}$) & (${\cal L}=100$ fb$^{-1}$) 
  & (${\cal L}=100$ fb$^{-1}$)  & (${\cal L}=100$ fb$^{-1}$) \\
\hline
$0.2$ & $1900$ & $1.5\times 10^{6}$ & $1.5\times 10^{7}$ & $1.2\times 10^{8}$\\
$0.3$ & $160$  & $2.3\times 10^{5}$ & $3.1\times 10^{6}$ & $2.8\times 10^{7}$\\
$0.4$ & $14$   & $5.5\times 10^{4}$ & $9.6\times 10^{5}$ & $9.7\times 10^{6}$\\
$0.5$ & $0.9$  & $1.7\times 10^{4}$ & $3.7\times 10^{5}$ & $4.3\times 10^{6}$\\
$0.6$ &  -     & $6400$             & $1.7\times 10^{5}$ & $2.1\times 10^{6}$\\
$0.8$ &  -     & $1200$             & $4.7\times 10^{4}$ & $7.1\times 10^{5}$\\
$1.0$ &  -     & $285$              & $1.6\times 10^{4}$ & $2.9\times 10^{5}$\\
$1.2$ &  -     & $81$               & $6800$             & $1.4\times 10^{5}$\\
$1.4$ &  -     & $26$               & $3100$             & $7.4\times 10^{4}$\\
$1.5$ &  -     & $15$               & $2200$             & $5.6\times 10^{4}$\\
$1.6$ &  -     & $8.8$              & $1600$             & $4.3\times 10^{4}$\\
$1.8$ &  -     & $3.1$              & $830$              & $2.6\times 10^{4}$\\
$2.0$ &  -     & $1.2$              & $470$              & $1.6\times 10^{4}$\\
$2.5$ &  -     & -                  & $130$              & $6100$\\
$3.0$ &  -     & -                  & $43$               & $2700$\\
$3.5$ &  -     & -                  & $15$               & $1300$\\
$4.0$ &  -     & -                  & $6.4$              & $690$\\
$5.0$ &  -     & -                  & $1.2$              & $230$\\
$6.0$ &  -     & -                  & -                  & $90$\\
$7.0$ &  -     & -                  & -                  & $40$\\
$8.0$ &  -     & -                  & -                  & $19$\\
$9.0$ &  -     & -                  & -                  & $9.7$\\
$10.0$ &  -    & -                  & -                  & $5.2$
\end{tabular}
\end{ruledtabular}
\end{table}


\end{document}